\documentclass[12 pt]{article}
\usepackage{graphicx}
\usepackage{amsmath}
\usepackage{latexsym}             
\usepackage{amsthm}
\usepackage{amssymb}
\usepackage{cite}
\usepackage{citesort}
\usepackage[dvips]{color}
\usepackage{setspace}
\begin{document}

\title{Efficiency, selectivity and robustness of nucleocytoplasmic transport\\}
\author{ A. Zilman $^1 $, S. Di Talia $^1 $, B. T. Chait $^3$ ,
 M. P. Rout $^2$, and  M. O. Magnasco $^1 $
\\ \emph{$^1 $ Laboratory of Mathematical Physics},\\
$ ^2$ \emph{Laboratory of Cellular and Structural Biology }\\ $^3 $\emph{Laboratory of Mass
Spectroscopy and Gaseous Ion Chemistry}\\\emph{ The Rockefeller University,
1230 York Ave, New York, NY 10021 }}
\date{} 
\maketitle  

\textbf{
Running title:} Physical model of transport through the NPC. 
\begin{abstract}
All materials enter or exit the cell nucleus through
nuclear pore complexes (NPCs), efficient transport 
devices that combine high selectivity and throughput. 
A central feature of this transport is the binding 
of cargo-carrying soluble transport factors to flexible, 
unstructured proteinaceous filaments called FG-nups that line 
the NPC. We have  modeled the dynamics 
of transport factors and their interaction with the flexible FG-nups as diffusion in an effective potential, using both analytical theory and computer simulations. We show that specific binding of transport factors to the FG-nups facilitates transport and provides the mechanism of selectivity. We show that the high
selectivity of transport can be accounted for by competition for both binding sites and space inside the NPC, which selects for transport factors over other macromolecules that interact only non-specifically with the NPC. We also show that transport is relatively insensitive to changes in the number and distribution of FG-nups in the NPC, due mainly to their flexibility; this accounts for recent experiments where up to half of the total mass of the NPC has been deleted, without abolishing the transport. Notably, we demonstrate that previously established physical and structural properties of the NPC can account for observed features of nucleocytoplasmic transport.  Finally, our results  suggest strategies for creation of artificial nano-molecular sorting devices.
\end{abstract}

\maketitle

\subsection*{Introduction}
The contents of the eukaryotic nucleus are separated from the cytoplasm by the nuclear envelope. Nuclear pore complexes (NPCs) are embedded in the nuclear envelope and are the sole means by which materials exchange across it. Water, ions, small macromolecules 
(smaller than $ 40$ kDa)\cite{macara_review} and small neutral particles (of diameter less than $ 5-6$ nm) can diffuse unaided across the NPC \cite{feldherr_gold}, while larger macromolecules (and even many small macromolecules) will generally only be transported efficiently if they display a signal
sequence, such as nuclear localization signals (NLSs) or nuclear
export signals (NESs), that bind to cognate soluble transport factors, which facilitate the passage through the NPC. The resulting transport factor-cargo complexes then traverse the NPC. The best studied transport receptors belong to a
family of structurally related proteins,
collectively termed $\beta $-karyopherins, although other transport
factors can also mediate nuclear transport,
particularly the export of mRNAs (reviewed in \cite{mike_review,macara_review,aebi_review,wente_review}). NPCs can pass cargos up to 30 nm diameter (such as mRNA particles) , at
rates as high as several hundred macromolecules per second, each one dwelling in the NPC for
only several milliseconds\cite{kubitschek_single,musser_single}. 

In this paper, we shall focus on karyopherin-mediated import, although our
conclusions pertain to other types of nucleocytoplasmic transport as well, including mRNA export. During import,
karyopherins bind cargoes in the cytoplasm via their NLSs. The
karyopherin-cargo complexes then translocate through NPCs to the nucleoplasm, where the cargo is released from the
karyopherin by a nuclear enzyme, RanGTP. The high affinity of 
RanGTP binding to karyopherins allows it to displace the cargoes 
from the karyopherins. Subsequently, karyopherins with bound RanGTP 
travel back through the NPC to the cytoplasm, where conversion of RanGTP to RanGDP is catalyzed by the cytoplasmic enzyme RanGAP. The energy released from GTP hydrolysis
is used to dissociate RanGDP from the karyopherins, which are now
ready for the next cycle of transport. Importantly, this GTP hydrolysis is the only  step in the process of nuclear import that requires input of metabolic energy.
Overall, the energy obtained from RanGTP hydrolysis is  used to create a concentration gradient of karyopherin-cargo complexes between the cytoplasm and the nucleus, and  the process of actual translocation across the NPC occurs purely by diffusion (reviewed in \cite{aebi_review,mike_review,macara_review,wente_review,stewart_review,mike2000}).

Conceptually, nuclear import can be divided into three stages:
first, the loading of cargo onto karyopherins in the
cytoplasm; then, the translocation of karyopherin-cargo
complexes through the NPC (cf. Fig. 1), and finally the release of
cargo inside the nucleus. The first and last stages have been
the subject of numerous studies, and are relatively well understood, being soluble-phase reactions amenable to biochemical
characterization (reviewed in \cite{aebi_review,mike_review,macara_review,stewart_review,wente_review}). The intermediate stage of transport is much less understood. Nevertheless, it is clear that  the ability of karyopherins (and other transport factors) to bind a particular class of NPC-associated proteins, known collectively as FG-nups, is a key feature of the transport process, and  allows them to selectively and efficiently pass with their cargoes through the NPC. In particular, experiments in which the FG-nup-binding sites on the karyopherins were mutated \cite{bayliss_stewart2002} show that disrupting the binding of karyopherins to FG-nups impairs transport. Current estimates of
the binding affinity of karyopherins to the FG-nups are in the range
10-1000 nM, (or $~5-10 k_B T $ per binding site) depending on the
FG-nup and karyopherin type\cite{rexach_affinity,importinbeta_nups_binding,nup1_binding}.

Transport through the NPC recently gave rise to several theoretical hypotheses \cite{mike2000,gorlich_gel,rabin,bruinsma,peters_traffic}. Here we develop a theory to explain the mechanism of the intermediate stage of nucleocytoplasmic transport - translocation through the NPC - which provides a physical basis for our virtual gating model \cite{mike_review,mike2000}.   

A theory of NPC-mediated transport has to answer several  major questions: 
\textit{i}) The transport of any macromolecules, even those potentially able to negotiate the NPC unaided, is increased by their specific association with NPCs \cite{Siebrasse_paper}. How does the NPC achieve  high transport efficiency of the cargoes of variable sizes and in both directions, through only passive diffusion of the transport factor-cargo complexes? \textit{ii)}  How does binding of transport factors to FG-nups facilitate transport efficiency  while maintaining a high throughput (up to hundreds of molecules per second per NPC) \cite{kubitschek_single,musser_single,mike_review}? 
\textit{iii)} NPCs largely exclude non-specific macromolecules in favor of transport factor-bound cargoes (e.g., \cite{macara_review}). How is this high degree of selectivity achieved? 
\textit{iv)} Neither deletion of up to half of the mass of the FG-nups, nor deletion of asymmetrically disposed FG-nups that potentially set up an affinity gradient, abolish transport \cite{wente_robustness}. Directionality of  transport  across the NPC  can even be reversed by reversing the concentration difference of the RanGTP \cite{weiss_direction_reversal}. How can we account for such a high degree of robustness? 

The model we present here provides answers to these questions, explaining the functional features of the NPC in terms of its known structural and physical properties. We show
how karyopherin binding to the flexible filaments inside the
NPC gives rise to  efficient transport. We show that the competition for
the limited space and binding sites within the NPC leads to highly
selective filtering. Finally, we show how the flexibility of the FG-nups
accounts for the high robustness of NPC-mediated transport with respect to structural changes\cite{wente_robustness}. We conclude by summarizing the salient features
of nuclear transport, and discuss verifiable experimental predictions of the model.

\subsection*{Setting up a physical model of NPC transport}

The NPC is a protein assembly spanning the nuclear envelope that contains a central channel,  about 35 nm in diameter, connecting the nucleoplasm with the cytoplasm. The
internal volume of this channel, as well as large fractions of the
nuclear and cytoplasmic surfaces of the NPC, are occupied by
unstructured polypeptide chains collectively known as FG-nups
because they carry unusually high numbers of repeats containing
phenylalanine-glycine pairs. FG-nups are known to be flexible
\cite{flexible_FG-nups,denning_flexible_FG,aebi_brush}. Since the FG-nups  also protrude 
 into the nucleus and the cytoplasm, the effective length of the nuclear pore complex is
estimated to be ~70 nm\cite{mike_review,aebi_review,macara_review}. 
The details of the distribution of the FG-nups inside the central channel and the external surfaces of the NPC, as
well as the exact number of binding sites on the FG-nups and on the
karyopherins, are not fully known. We shall therefore make no specific
assumptions about the distribution of
FG-nups, interactions between them, and their density, degree of flexibility or conformation within the NPC.

We represent the transport through the NPC as a combination of two independent processes contributing to the movement of the
karyopherin-cargo complexes through the central channel of the NPC: \textit{(i)} the
binding and unbinding of the karyopherins to the
FG-nups, and textit{(ii)} the spatial diffusion of the complexes,
either in the unbound state, or while still bound to a flexible FG-nup\cite{aebi_review}. 
The complexes entering the NPC
from the cytoplasm stochastically hop back and forth inside the
channel until they either reach the nuclear side, where the cargo is
released by RanGTP, or return  to the cytoplasm. Detachment from the FG-nups and exit from the NPC can be either thermally activated, or catalyzed by RanGTP directly at the nuclear exit of the NPC \cite{macara_review}. A schematic illustration of transport through the NPC is shown in
Fig. 1.

\subsection*{Transport efficiency arises from the karyopherins' ability to bind
to FG-nups.}

It is important to distinguish between two different properties of
the transport process: the \textit{speed}  and   the \textit{probability} with which individual complexes that enter from the cytoplasm traverse the NPC to nuclear side. \cite{gardiner,berezhkovskii_ptr,schulten_glycerol}. As we show
below, binding of karyopherins to the FG-nups increases their transport efficiency, i.e., the probability of them traversing the NPC; in the absence of such binding, the probability of traversing the NPC is low.

For simplicity, we shall assume that the unbinding and rebinding occur faster than lateral diffusion of karyopherin-cargo complexes along the channel(our conclusions nevertheless were verified by computer simulations for any ratio of binding-unbinding rate to diffusion rate). In this limit, the hopping between
discrete binding sites can be approximated by diffusion in an
effective potential which combines two effects: entropic  repulsion of the FG-nups, as the large
cargoes have to compress and displace them to enter the channel,
and attraction due to binding to the binding sites in the FG-nups \cite{mike_review,mike2000,aebi_brush}; this approximation is justified below.
Thus, we represent the transport of karyopherin-cargo complexes
through the NPC as a one-dimensional diffusion in a potential $U(x) $ (expressed in units of $k_BT $), in the
interval $0<x<L$, as illustrated in Fig. 2.  We solve the model for an arbitrary shape 
of the potential, determined by the distribution of the binding sites on the FG-nups
along the channel.  We do not directly model the
diffusion of complexes outside of the NPC. Instead, we assume that
karyopherin-cargo complexes stochastically enter the NPC from the
cytoplasm, with an average rate $J$ at $x=R$; $J$ is
proportional to the concentration of the karyopherin-cargo complexes
in the cytoplasm \cite{berg_book}.  The release of karyopherin-cargo complexes from FG-nups by RanGTP at the nuclear exit is modeled by imposing an exit flux $J_e $  at $x=L-R$, as shown in Fig. 2.  Thus, the length of the NPC  corresponds to the interval from $x=R$ to $x=L-R$; the regions of length $R$ (of the order of the  width of the channel \cite{berg_book}) at both ends of the interval correspond to the distance outside the NPC over which the particles diffuse away by three-dimensional diffusion into either the
nucleoplasm or the cytoplasm. We use absorbing boundary conditions at $x=0 $ and $x=L $ that
correspond to a karyopherin-cargo complex returning back to the cytoplasm, or going through to the nucleus, respectively. 

We neglect the
difference in the diffusion coefficient of the complexes inside
and outside the NPC in order to focus on the role of
karyopherin binding in the import process. We also assume that no current
enters from the nucleus, as the cargoes are released from the
karyopherins in the nucleus by RanGTP. Finally, we neglect 
 the variations of the potential in the direction perpendicular to the channel axis. 
The effect of these factors will be studied in
detail elsewhere. Under the
above assumptions, transport of the karyopherin-cargo complexes
through the NPC is described by the diffusion  equation for the
 density of the complexes inside the channel, $\rho(x)$ \cite{gardiner}:
\begin{equation}\label{smoluchowski1}
 \frac{\partial \rho(x)}{\partial t}=-\frac{\partial J(x)}{\partial x}
\end{equation}
where the local current $J(x) $ is given by:
\begin{equation}\label{smoluchowski2}
J(x)=-D e^{-U(x)} \frac{\partial}{\partial x}[ e^{U(x)}\rho(x) ]
\end{equation}
where $D$ is the diffusion coefficient.

The entrance current $J $ splits
into $J_0 $, and $J_M $, corresponding to the flux of complexes
returning to the cytoplasm and going through to the nucleus,
respectively. At a position $x=L-R $, the
transmitted flux $J_M $ splits into  $J_e $ and $J_L$, which
correspond  to the karyopherin-cargo complexes released from the
FG-nups by RanGTP and to thermally activated release, respectively (cf. Fig. 2).

The steady state solution of
eqs.(\ref{smoluchowski1},\ref{smoluchowski2}) with entrance flux $J$ and satisfying the
boundary conditions $P(0)=P(L)=0 $, is
\begin{eqnarray}\label{eqn_P}
\rho(x)& = & |J_0|\frac{1}{D}e^{-U(x)}\int_0^{x} e^{U(x')}dx' \;\;\;\;\text{for}\; 0<x<R\nonumber\\
\nonumber\\
\rho(x)& = & \frac{1}{D}e^{-U(x)}\left[|J_0| R-J_M\int_{R}^{x}dx' e^{U(x')}\right]\;\;\;\\&& \text{for}\;
R<x<L-R\nonumber\\
\nonumber\\
\rho(x)& = & J_L\frac{1}{D}e^{-U(x)}\int_{x}^{L}e^{U(x')}dx'\;\; \text{for}\; L-R<x<L\nonumber
\end{eqnarray}
The sum of the flux of karyopherin-cargo complexes  going through the NPC and of those returning to the cytoplasm is
equal to the total flux of complexes  entering the NPC; hence $|J_0|+J_M=J $; similarly, 
$J_M-J_L=J_e $. The flux $J_e $ is proportional to the number  of complexes present at the
nuclear exit, and to the frequency $J_{\text{ran}} $ with which  RanGTP molecules hit the nuclear exit of the NPC: $J_e=J_{\text{ran}}\rho(x=L-R)R $. Recalling that that  the potential outside the channel is  zero ($U(x)=0 $) for $0<x<R $ and $L-R<x<L $, and using the continuity of $\rho(x) $
at $x=L-R$, one obtains for $P_{tr}$, the probability of a given karyopherin-cargo complex  reaching the nucleus: 
\begin{equation}\label{ptr_ran}
 P_{tr}=J_M/J =\frac{1}{2- K/(1+K)+\frac{1}{R}\int_{R}^{L-R}dxe^{U(x)}}
\end{equation}
where $K= (J_{ran}R^2/D )\;e^{-U(L-R)} $.

Eq.(\ref{ptr_ran}) is the main result of this section and has
several important consequences. The probability $P_{tr} $ defines the transport
efficiency, which is seen to increase with the potential depth $E$, (defined as $E=-\min_x U(x) $, Fig. 2), proportional to the binding strength of the karyopherins to the FG-nups. In the absence of binding, $P_{tr}$ is small ($\sim R/L $), so that a complex will on average return  to the cytoplasm 
soon after entering the NPC. An attractive potential inside the NPC \textit{increases} 
the time the complex spends inside the NPC and thus \textit{increases} 
the probability that it reaches the nuclear side, rather than returns to the  cytoplasm.

When RanGTP only releases cargo from its karyopherin, but not from the FG-nups
(i.e. $J_{\text{e}}=0 $), the
maximal translocation probability $P_{tr}$ is 1/2. However, in the case when RanGTP also releases
karyopherin-cargo complexes from FG-nups, the translocation probability $P_{tr}$ can reach 1.
Importantly, the latter effect is more pronounced for large $K$, that is for strong binding at
the exit.

The second important consequence of eq. (\ref{ptr_ran}) is that the translocation probability $P_{tr}$ 
depends only weakly on the \textit{shape} of the potential $U(x) $. Thus, we predict that the
transport properties of the NPC are relatively insensitive to details of the distribution
of FG-nups inside the NPC, and to the distribution of the binding sites on FG-nups.

\subsection*{Limitations of space and number of binding sites provide a mechanism for            selectivity}\label{sec_jamming}
The discussion  of the previous section neglects the mutual interactions between karyopherin-cargo complexes. 
However,  a large interaction strength $E$ increases transport efficiency  at the expense of an increased transport time $T(E)$ (which grows roughly exponentially with $E$ \cite{gardiner}), leading to accumulation of particles inside the channel.
Therefore, for high $E$, interactions between the complexes due to steric occlusion and  decrease in the  available binding sites become important.

To quantitatively investigate how mutual interference between
translocating karyopherin-cargo complexes affects transport efficiency, we performed dynamic Monte Carlo
simulations of the  diffusion of complexes inside the NPC, in the potential $U(x)$, using a variant of the Gillespie algorithm \cite{lebowitz,gillespie,ledoussal}. The simulations are a discrete version of the continuum formulation of the previous section. The interval
$[0,L] $ is represented by $N $ discrete "sites"; these sites do not represent the actual binding sites, but correspond to the length of a diffusion step. For simplicity, we allow only one particle at each site at any moment of time, which models mutual occlusion between complexes; however, our results pertain for any allowed occupancy.  In line with our analytical model above, karyopherin-cargo complexes are deposited at the site $i_R$, if it is unoccupied,
with a probability $J\frac{L^2}{DN^2}$ per simulation step. When a complex reaches site $i=0 $  (cytoplasm) or $i=N$ (nucleus), it is removed.
The complexes present at site $i=N-i_R$ can  be removed directly, with the probability $J_{\text{ran}}\frac{L^2}{DN^2} $, which models the effect of the
release of the complexes from FG-nups by nuclear RanGTP. Once inside the channel,  a complex present at site $i$ can hop  to an adjacent unoccupied site
$i\pm 1 $ with the following probability:
\begin{equation}\label{jumping_probability}
P(i\rightarrow  i\pm 1)=r_{i,i\pm 1}/(J+J_{ran}x_{N-i_R}+\sum_{i=1}^N r_{i,i\pm 1})
\end{equation} where  $x_i $ is the site occupancy: $x_i=0 $ if the
site is unoccupied, and $x_i=1 $ if a particle is present at the site; $r_{i,i\pm 1}=\frac{D}{(L/N)^2}\exp((U_{i}-U_{i\pm 1})/2)x_i(1-x_{i\pm 1}) $
is the transition rate from site $i$ to $i\pm 1$\cite{lebowitz,gillespie,ledoussal}. 

The results of our simulations are shown in Fig. 3.  For low interaction strength  $E$ the translocation probability curves for all entrance fluxes $J $ collapse onto a single line
which is predicted by the analytical solution of the previous section, because in this regime there are few complexes simultaneously present in the channel, and the interactions between them are negligible. For stronger binding, karyopherin-cargo complexes accumulate inside the channel, blocking the inflow of additional complexes.

The main conclusion of the simulations, as shown in Fig.3 \textbf{A}, is that the transport  through the NPC is maximal for an optimal value of the interaction strength $E_c $, which depends on the entrance rate $J$. The decrease in NPC throughput at high $E $  becomes significant when the number of karyopherin-cargo complexes in the channel reaches a certain critical value, $M_c$. This critical occupancy of the channel $M_c$ is proportional both to the entrance flux $J$, and to the residence time $T(E_c)$, $M_c=J T(E_c) $, and therefore  the optimal interaction strength is higher for low entrance fluxes as illustrated in panel \textbf{B} of Fig. 3 \ref{fig_trans_prob_ran}.

Existence of an optimal interaction strength provides a mechanism for the selectivity of
NPC-mediated transport. The translocation probability is high for a particular strength of interaction of
karyopherins with FG-nups, while macromolecules that do not interact with FG-nups
are filtered out.

\subsection*{Competition between non-specifically binding macromolecules and 
karyopherins enhances the selectivity of the NPC}
As shown in the previous section, the binding of karyopherins to FG-nups
provides a mechanism of selectivity, because  transport of the karyopherin-cargo complexes is highest for a specific value of their interaction strength with the  FG-nups. However, the maximum depicted in Fig. 3\textbf{A} is broad; the translocation probability is significant even for binding strengths
considerably lower than the optimal one. For instance, if the optimal interaction strength is
$E=10  kT$, macromolecules whose interaction strength is $5 kT$ a have probability of reaching  the nucleus that is just a half of the optimal one.  This broad maximum allows NPC-mediated import to function efficiently across a broad range of transport factor binding strengths. On the other hand, this might also permit passage of macromolecules that bind
non-specifically to FG-nups (e.g., due to electrostatic interactions). However, proper functioning of living cells requires a high selectivity of the NPC - how might this be achieved? 

So far, Fig. 3 only takes into account the competition between  complexes of identical binding strength for space inside the channel. However, in a situation where optimally-binding karyopherins compete for space and binding sites inside the channel with weakly-binding macromolecules, the passage of the latter is sharply reduced which sharply increases the selectivity of the NPC. Qualitatively, because the residence time is
higher for  strongly-binding karyopherins, a weakly-binding macromolecule upon entering the channel
will with high probability find it blocked by a karyopherin. 
Thus, the low affinity macromolecule will with high probability return to the cytoplasm due to a relatively low residence time. On the other hand, if a karyopherin enters a channel that is already   occupied by another karyopherin, there is a high probability that it will reside inside the channel long enough for the first karyopherin to get through. As free karyopherins exchange back and forth across the NPC constantly, there will always be karyopherins (or other transport factors) binding in the NPC and so excluding non-specific macromolecules, making the NPC a remarkably efficient filter. 

These heuristic arguments were verified via computer simulation. Two species of particles  of different binding affinities (representing the karyopherins and another macromolecule that can non-specifically bind to the FG-nups), are deposited stochastically at the NPC
entrance with the same average rate $J$. As in the previous section, the particles diffuse inside the channel until they either reach the nucleus, or return to the cytoplasm; a site can be occupied by only one particle.

The results of the simulations are presented in Fig. 4, which compares the
transport efficiency of strongly binding macromolecules to that of weakly binding ones in the case when they compete for the space inside the NPC.
The competition for the space inside the channel between the translocating particles dramatically narrows
the selectivity curve as compared with Fig. 3.
This effect is a novel mechanism for the enhancement of transport selectivity beyond what is
expected from the binding affinity differences alone. In contrast to other mechanisms of specificity (e.g. kinetic proofreading\cite{proofreading}), no additional metabolic energy is required for this enhanced discrimination. Instead, selectivity is achieved by competition producing a differential NPC response to two ranges of binding affinities.  In the range of higher binding affinities occupied by transport factors, passage across the NPC is efficient; whereas in the low range of affinities, transmission is effectively prevented.

\subsection*{The flexibility of FG-nups accounts for the robustness of NPC-mediated transport}
In the previous sections, we have used a continuous potential to model transport through the NPC. However, in reality the translocating karyopherin-cargo complexes hop between discrete binding sites that are located on the flexible FG-nups, which fluctuate in space around their anchor point due to thermal motion  \cite{flexible_FG-nups,aebi_brush}. This allows complexes to diffuse along the channel while remaining bound to an FG-nup. A complex can also unbind from an FG-nup and rebind again to the same or a neighboring FG-nup, moving while unbound by passive diffusion. In this section, we elucidate how the \textit{number} of FG-nups inside the channel affects transport.

The translocation of the karyopherin-cargo complexes through the NPCcan be described as a diffusion in an array of potentials, as illustrated in Fig. 5, where each potential well  $U_i(x) $ represents an FG-nup. The shape of each well depends on  the binding strength of the karyopherin-cargo complex and the rigidity and the length of an FG-nup, which determine the cost of its entropic stretching in the process of spatial diffusion \cite{aebi_brush}. The blue line corresponds to the unbound state. Although for the purposes of illustration all the wells are shown to have the same form, the subsequent results are valid for arbitrary distribution of potential shapes. We shall denote the density of the karyopherin-cargo complexes in the $i$-th well as $\rho_i(x) $ and the density of unbound complexes as $\rho_0(x) $.  The lateral diffusion of the complexes, combined with the binding and unbinding to the FG-nups is then described by the following equations \cite{julicher_rev_mod}:

\begin{eqnarray}\label{eq_crossing_wells}
\frac{\partial \rho_0(x)}{\partial t}&=&\frac{\partial}{\partial
x}e^{-U_0(x)}\frac{\partial}{\partial x}e^{U_0(x)}\rho_0(x)+\nonumber\\
&+&\sum_i[r_{i0}(x)\rho_{i}-r_{0i}(x)\rho_0(x)]\nonumber\\
\frac{\partial \rho_i(x)}{\partial t}&=&\frac{\partial}{\partial
x}e^{-U_i(x)}\frac{\partial}{\partial x}e^{U_i(x)}\rho_i(x)+\nonumber\\
&+&r_{0i}(x)\rho_{0i}-r_{i0}(x)\rho_i(x)
\end{eqnarray}
where $r_{i0}(x)$ and $r_{0i}(x)$ are the local unbinding and binding rates, respectively. They are related by the detailed balance condition: $
r_{0i}(x)/r_{i0}(x)=e^{-U_i(x)+U_0(x)}$. If the unbinding rates are fast  compared to the diffusion time
across the wells (i.e. $r_{i0}^{-1}e^{U_0-U_i}\gg D/(2 p^2) $), the relative densities of bound and unbound complexes are at their local equilibrium Boltzmann ratio \cite{julicher_rev_mod}:
$\rho_i(x)=\frac{e^{-U_i(x)}}{\sum_i e^{-U_i(x)}}\rho(x)$ where $\rho(x)=\sum_i \rho_i(x) $ is the total density of the complexes at a position $x$. 
Summing up the equations (\ref{eq_crossing_wells}), one obtains an equation for the total complex density $\rho(x) $:

\begin{eqnarray}
\frac{\partial \rho(x)}{\partial t}=\frac{\partial}{\partial
x}e^{-U_{\text{eff}}(x)}\frac{\partial}{\partial x}e^{U_{\text{eff}}(x)}\rho(x)
\end{eqnarray}
where $U_{\text{eff}}=-\ln(e^{-U_0(x)}+\sum_i e^{-U_i(x)})$  \cite{julicher_rev_mod}. Thus, the process of translocation through the NPC can be described as simple lateral diffusion in the effective potential $U_{\text{eff}} $, which leads to the eqs. (\ref{smoluchowski1},\ref{smoluchowski2}).

As we have shown in previous sections, the transport properties of the NPC are relatively insensitive to the detailed shape of the effective potential. Thus, we predict that the transport trough the nuclear pore is robust with respect to the variations in the number of the FG-nups, as indeed observed experimentally \cite{wente_robustness}.

Even more strikingly, the transport properties of the NPC are not sensitive to  the number of the FG-nups also in the case when the FG-nups are distributed sparsely inside the NPC, without a significant overlap.
We illustrate this issue through a marginal  case where the fluctuation regions of each FG-nup barely touch, represented by the potential shown by the black
line in Fig. 6\textbf{A}. The flat central  part of each potential well corresponds to
the diffusion of the karyopherin-cargo complex while  bound to an FG-nup, and the sharply rising
regions at the borders correspond to unbinding of the complex and its transfer to the next filament. Narrow wells correspond to relatively rigid
filaments, while wide wells correspond to
flexible filaments that can stretch to a long distance without significant entropic cost.  The potential wells can have different 
widths $p_i$ so that $\sum_{i=1}^np_i=L-2R $. All the potential wells
have the same shape, $U_0 $, re-scaled to the width of an individual well,  so that  the
potential  at a  point $x=\sum_{j<i}p_j+\Delta x $ is:
$ U(x)=U_0(\Delta x/p_i)$.               

Crucially, the transport properties of the potential shown in Fig.
6\textbf{A}  \textit{do not depend on the number of wells}. Both the translocation probability and the residence time are equivalent for the
multi-well potential shown in black, and the single well potential shown in red, obtained by re-scaling an individual black well to the whole length $L-2R$ of the channel. Indeed, it follows from eq. (\ref{ptr_ran}) that  the translocation probability for the
multi-well potential with $n$ wells, is
\begin{eqnarray}
P_{tr}&=&\frac{1}{2-K/(1+K)+\frac{1}{R}\sum_{i=1}^{i=n-1}\int_{0}^{p_i}e^{U_0(x/p_i)}dx}\nonumber\\
&=& \frac{1}{2-K/(1+K)+\frac{\sum_i p_i}{R}\int_{0}^{1}e^{U_0(y)}dy}
\end{eqnarray}
which is independent of the number and the width of the wells because $\sum_i p_i=L-2R $. We prove in the Supporting Information that the residence time is similarly independent of the number of
wells.  Since both translocation probability and residence time are independent of
the number of wells, the transport properties do not depend on  the number of wells even for
high entrance rate or binding strength, when jamming is important, as verified by 
computer simulation (Fig. 6\textbf{B}).

This result highlights the robustness of our model of NPC transport; in multiwell potentials of this type, the NPC's transport properties do not depend on the specific number of FG-nups, so long as they are flexible enough for their fluctuation regions to overlap, permitting complexes to freely transfer from one filament to the next. 

\subsection*{Discussion}

We have presented a physical model of transport through the NPC. The physics of diffusion in a channel and binding to flexible filaments accounts for the major observed features of NPC transport, without the need for further detailed assumptions such as the conformation and distribution of FG-nups inside the channel or the shape of the channel itself. 

A crucial component of our model is the interaction between karyopherins and FG-nups, which provides a mechanism for both efficiency and selectivity. We show that in the absence of binding, the probability of crossing the NPC is much lower than the probability of returning to the cytoplasm.
The binding of karyopherins to FG-nups increases the residence time of the karyopherin-cargo complexes within the NPC, preventing their premature return to the cytoplasm, and hence increasing their chance of diffusing to the nucleus, i.e., increasing the efficiency of their transport.

We show that the selectivity of the NPC arises from a balance between the \textit{probability} and the \textit{speed} of transport of individual karyopherin-cargo complexes. Similar ideas have been suggested to account for the transport properties of ion channels and porins\cite{berezkovskii_jamming,berezhkovskii_ptr,gardiner,schulten_glycerol}. In our model, the probability of a karyopherin-cargo complex reaching the nucleus increases with the binding strength to FG-nups, but at the expense of increasing residence time inside the NPC; eventually, complexes spend so much time in the NPC that they impede the passage of other complexes through channel. Therefore, there is an optimal value of the binding strength of karyopherins to FG-nups that maximizes their transport efficiency through the NPC. This optimal binding strength is higher for low entrance fluxes $J$, because at low fluxes the accumulating complexes can reside longer in the channel without blocking it (Fig. 3).  This observation could account for the proliferation of different karyopherins types; the binding strength of each karyopherin type might be related to the total relative abundance of its cargoes. Our model applies to both export and import processes, and explains how highly efficient nucleocytoplasmic transport can be achieved in both directions by pure diffusion. 

For proper cell functioning, the NPC should selectively transport karyopherin-cargo complexes, effectively filtering out any macromolecules that do not bind specifically to the FG-nups. However, the basis for discrimination between specifically and non-specifically binding macromolecules may be as little as a few  $k_B T $. 
We propose a novel mechanism that further enhances the specificity of NPC transport. This mechanism relies on the direct competition between karyopherins and non-specifically binding macromolecules; they compete for space and binding sites in the channel. As a consequence of their stronger binding, karyopherins have a longer residence time within the channel as compared with non-specifically binding macromolecules which are therefore out-competed for space and binding sites within the channel. The constant flux of cargo bound- or free karyopherins between the nucleus and cytoplasm therefore effectively excludes non-specifically binding macromolecules from the channel. Remarkably, although no metabolic energy is expended in this filtering process\cite{proofreading}, the resulting selectivity is much higher that might be expected from only the binding affinities differences between specific and non-specific macromolecules (Fig. 4).

We have also shown that transport efficiency is enhanced when RanGTP directly releases karyopherins from their binding sites on FG-nups\cite{rexach_affinity,macara_review} at the NPC exit - an enhancement that increases with the binding  strength at the nuclear exit. Physically, high affinity binding sites at the nuclear exit of the NPC decrease the probability of return, once a complex has reached the nuclear side.
This prediction may account for the observed high affinity binding sites that are localized at the nuclear side on the NPC in import pathways and cytoplasmic side in export pathways (reviewed in \cite{mike_review}).

Although the transport properties of the NPC depend strongly on the
magnitude of the interaction strength, we predict that transport depends 
only weakly on spatial variations of the  binding sites along the channel. 
Most importantly, the \textit{number} of flexible FG-nups inside the NPC does not significantly affect transport, as long as their fluctuation regions can overlap (Figs. 5 and 6). This prediction could account for recent experiments in which up to half the total mass of FG-nup flexible regions in NPCs were deleted without seriously hampering nucleocytoplasmic transport \cite{wente_robustness}. Moreover, we predict that a gradient of binding affinity across the NPC does not, by itself, increase the throughput compared to a uniform distribution of the same sites, explaining how transport can even be reversed across the NPCby reversing the gradient of RanGTP \cite{weiss_direction_reversal}. Only in combination with a high affinity trap at the exit, and high Ran activity in releasing the karyopherins from this trap \cite{macara_review}  can the affinity gradient improve the throughput through the NPC \cite {mike2000,mike_review}.

Direct tests for our model's predictions can now be made by experimentally varying the effective potential experienced by the karyopherin-cargo complexes inside the NPC, by systematically introducing mutations into the binding sites\cite{bayliss_stewart2002}, changing the cargo size\cite{feldherr_gold}, or using cells with genetically modified numbers of the FG-nups\cite{wente_robustness}. Finally, any device built according to the principles outlined above would possess the transport properties described by our model, suggesting strategies for creation of artificial nano-molecular sieves.

\subsection*{Materials and methods}
The simulations were written in C language and run on  a cluster of UNIX processors. The simulation algorithm is described in the text. Analytical calculations were partially performed using Mathematica 5.1 package.

\textbf{Acknowledgements}
The authors are thankful to J. Aitchison, S. Bohn, T. Chou, R. Peters, B. Timney, J. Novatt and G. Stolovitzky for helpful comments. This work was supported by  NIH grants RR00862 (B.T.C.), GM062427 (M.P.R.), GM071329 (M.P.R. B.T.C, A.G.Z, M.O.M.) and RR022220 (M.P.R., B.T.C.)


\newpage
\begin{figure}[htbp]
\includegraphics[width= 15 cm]{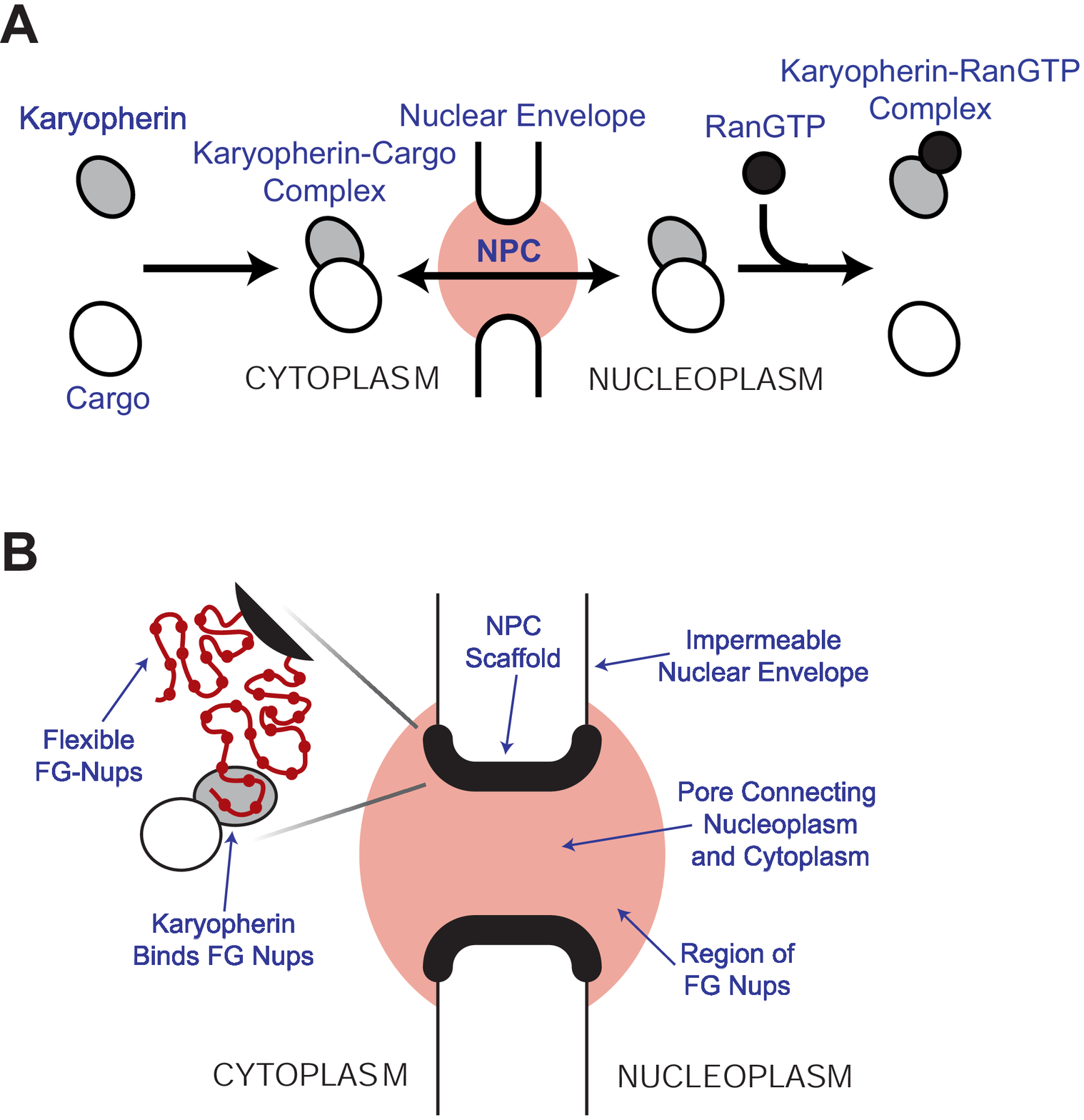}
\caption{\textbf{Main features of NPC and of the nuclear import.} \textbf{A}: Schematic illustration of the nuclear import process. The karyopherins bind the cargo in  the cytoplasm and transport it to the nucleus, where the cargo is released by RanGTP. \textbf{B}: Diagram of the NPC. The nucleus and the cytoplasm are connected by a channel, which is filled with flexible filaments, FG-nups. The karyopherins carrying a cargo  enter from
the cytoplasm and hop between the binding sites on the fluctuating FG-nups until they either
reach the nuclear side of the NPC, or return to the cytoplasm. }\label{brush_cartoon}
\end{figure}

\begin{figure}[htbp]
\includegraphics[width= 15 cm]{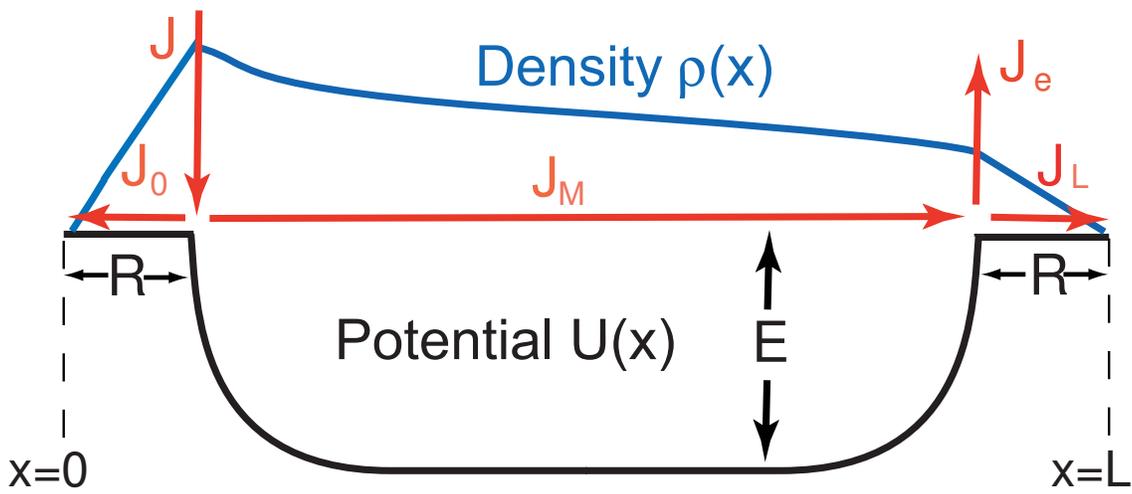}
\caption{\textbf{Transport through the NPC is modeled as diffusion in an energy landscape.} The NPC channel is represented by a potential well $U(x) $, shown in black. The complexes enter the NPC at $x=R$ at an average rate  $J$. A fraction of the entrance flux, $J_M$, goes through to the nucleus. The rest return to the cytoplasm at an average rate $J_0 $. The  exit of the complexes from the channel occurs either due thermal activation, with the rate $J_L$, or by activated release by RanGTP, with the rate $J_e $. Steady state particle density inside the channel $\rho(x) $ is shown in blue.  }\label{brush_cartoon}
\end{figure}

\begin{figure}[tbp]\label{fig_trans_prob_ran}
\includegraphics[width= 13 cm]{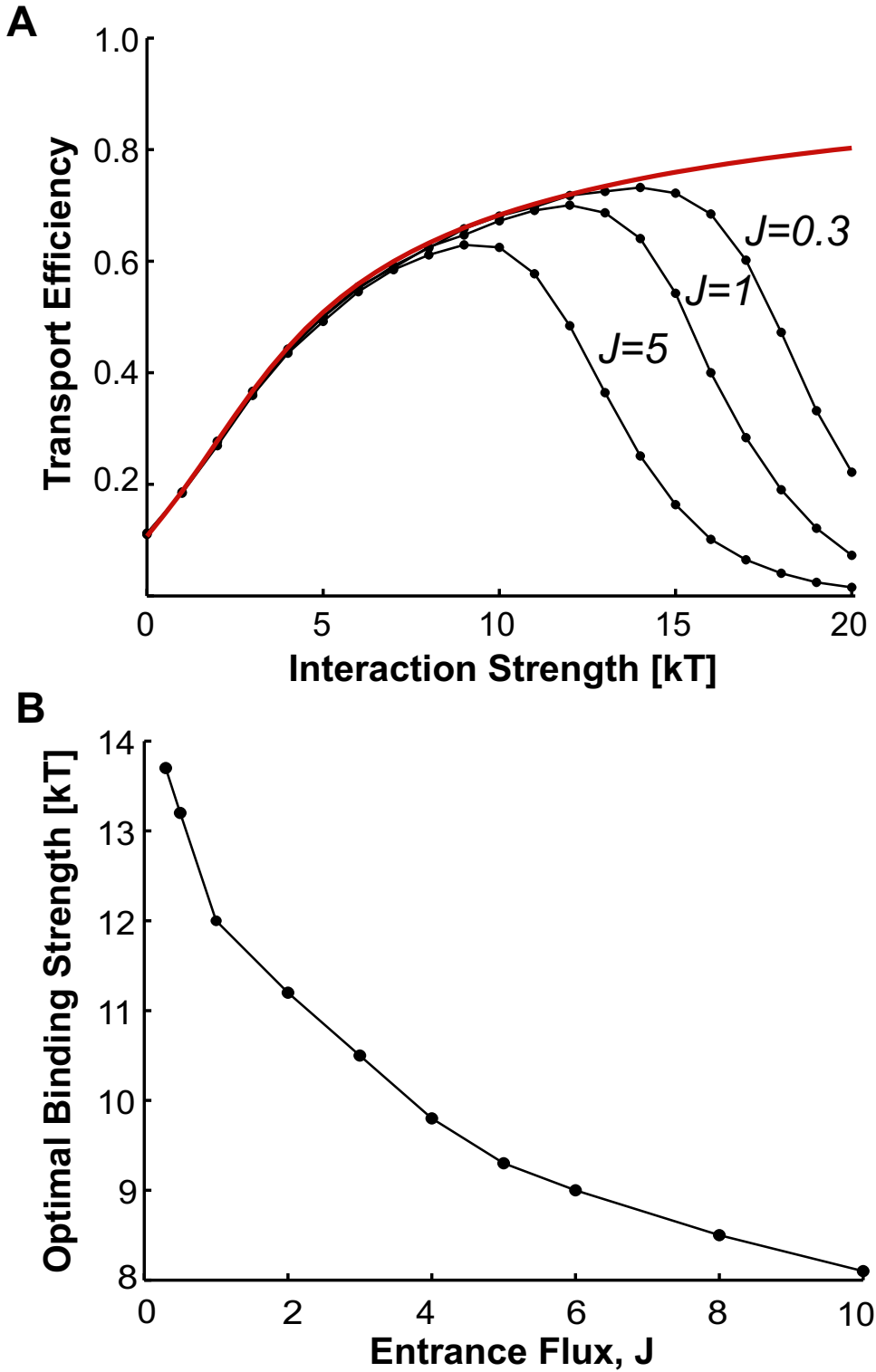}
\caption{\textbf{Transport efficiency is determined by the interaction strength.} \textbf{A}: Transport efficiency, as given by  probability to reach the nucleus is shown as a
function of the interaction strength $E $. (RanGTP activity in the nucleus is represented by $J_{\text{ran}}L^2/(N^2D)=1.5$). The curves correspond to three different values of the entrance rate $J$ (measured in units of $10^{-4}R^2/(16 D))$; the red line is the low rate limit of eq.(4). For any entrance rate, the transport efficiency is maximal at a specific value of the interaction strength, which provides a mechanism of selectivity. \textbf{B}: Optimal interaction strength of panel \textbf {A} as a function of the incoming rate $J $ (in units of $10^{-4}R^2/(16 D) $), for $ J_{\text{ran}}L^2/(N^2D)= 1.5$. Black dots are simulation results.}
\end{figure}

\begin{center}
\begin{figure}[htbp]\label{fig_competition}
\includegraphics[width= 15 cm]{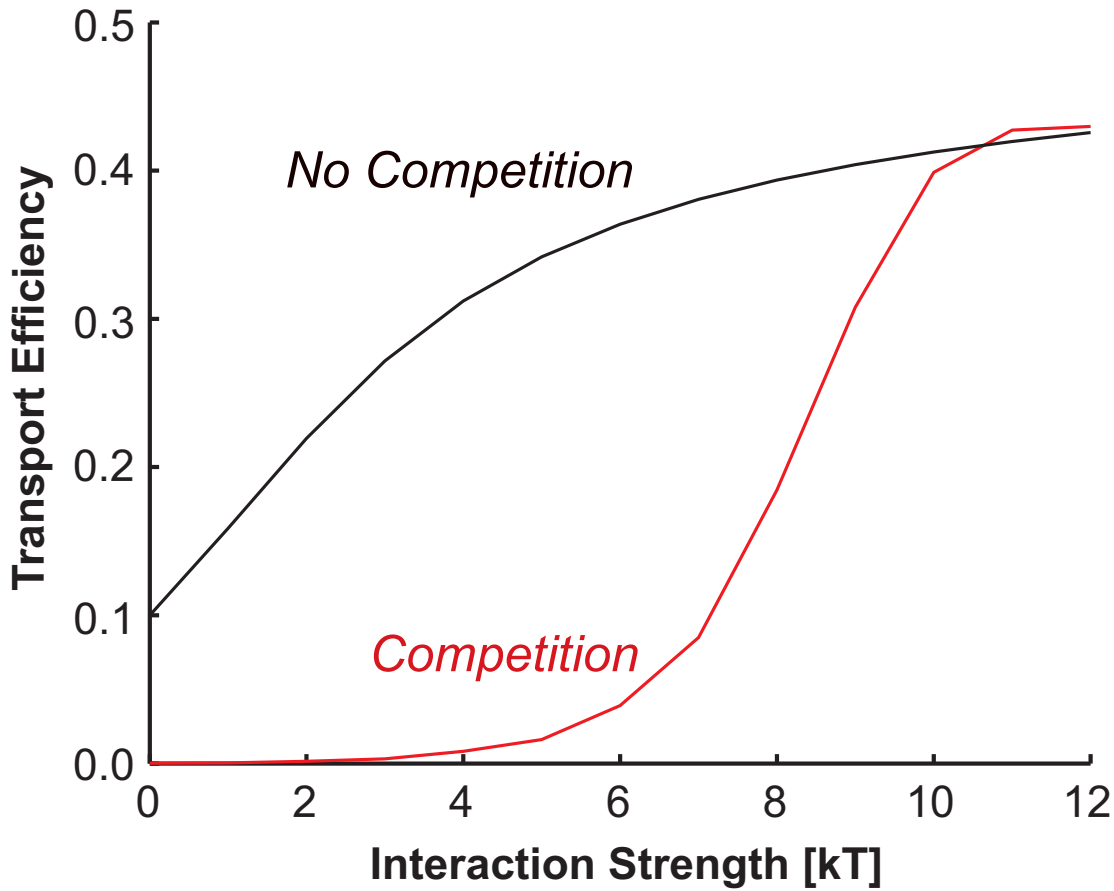}
\caption{\textbf{Karyopherins efficiently exclude non-specifically binding macromolecules  from the NPC.} Black line: translocation probability of the complexes as a function
of the interaction strength for a single species. Red line: translocation probability of the weakly-binding species in an equal mixture of weakly- and strongly- binding species, as a function of the interaction strength of the weakly-binding species; the interaction strength for the strongly binding species is $12 k_B T $. Translocation of the weakly binding species is sharply reduced in the presence of the strongly binding species.}
\end{figure}
\end{center}

\begin{figure}[htbp]\label{fig_crossing_wells}
\includegraphics[width= 15 cm]{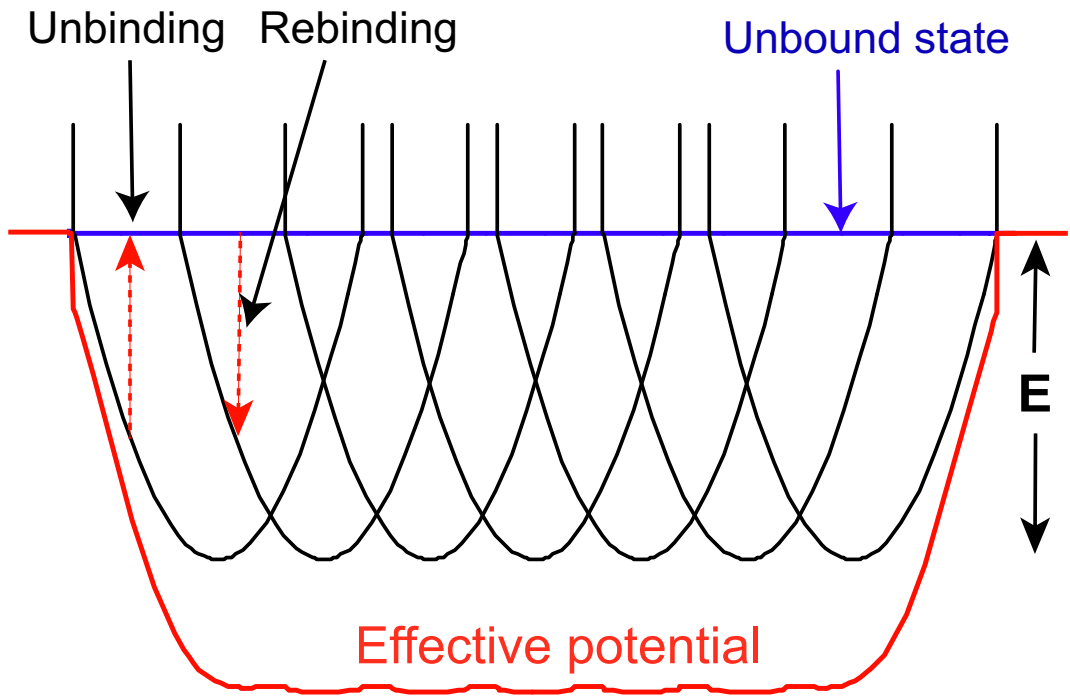}
\caption{\textbf{Discrete overlapping FG-nups can be approximated by a smooth effective potential.} Transport through the NPC can be represented as diffusion in an array of potential wells that represent flexible FG-nups whose fluctuation regions overlap.
The red dotted lines correspond to the complexes unbinding from and rebinding to the FG-nups.
Blue line is the unbound state. Red line shows the equivalent potential in the case when the unbinding of the complexes from the FG-nups is much faster than the lateral diffusion across an individual well.}
\end{figure}

\begin{figure}\label{fig_effective_potential}
\includegraphics[width= 15 cm]{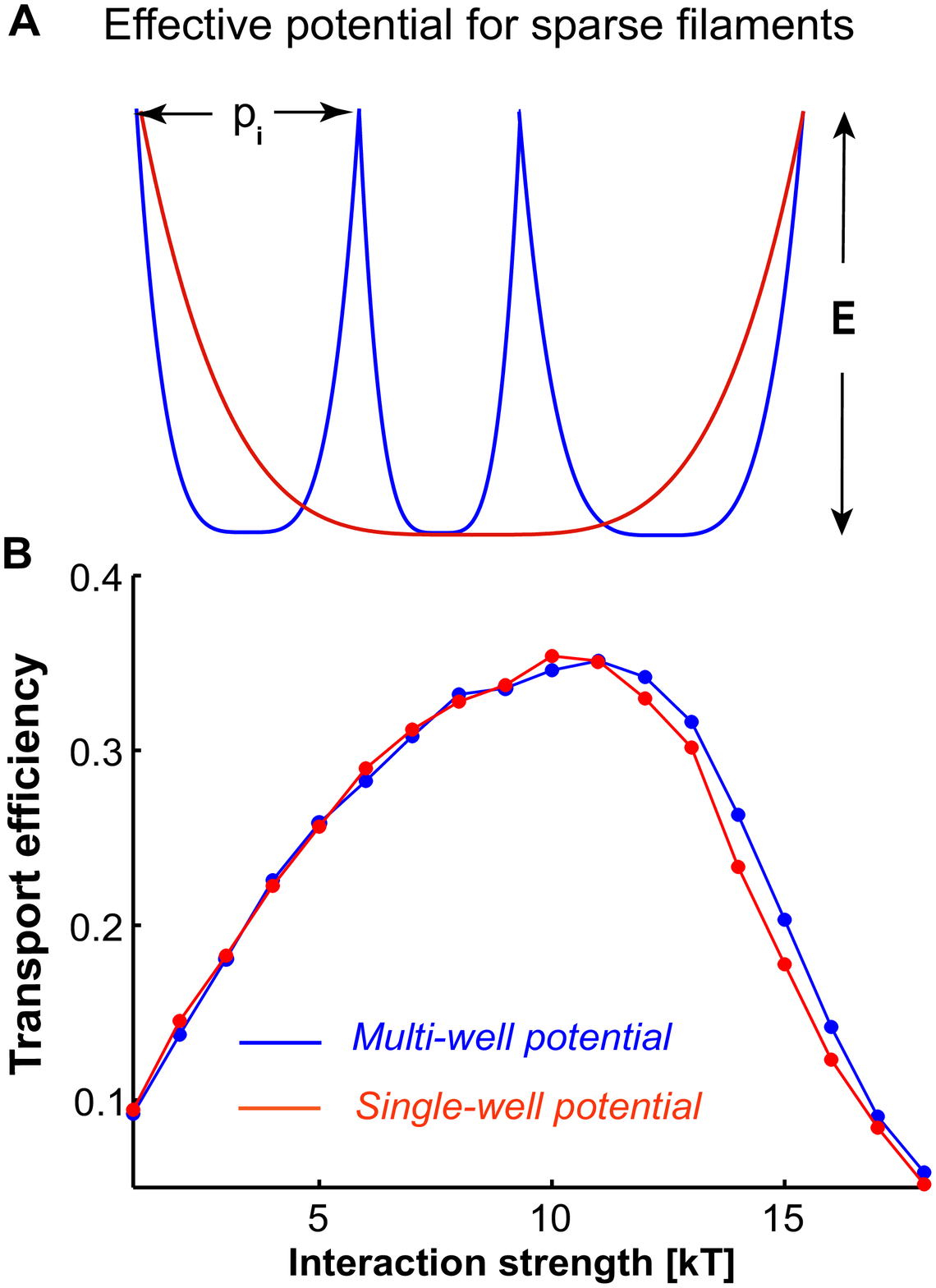}
\caption{\textbf{The number of the FG-nups does not significantly affect the transport properties.} \textbf{A}: Effective potential for sparse flexible FG-nups is shown in blue line. Each well corresponds to an FG-nup. The transport properties in this multi-well potential are \textit{independent} of the number of wells, and hence equivalent to those in the single-well potential, shown in red line.  
\textbf{B}: Numerical simulations show essentially identical transport efficiencies in a multi-well potential (blue line) and in the single-well potential (red line); $U_0(x)=((x/p)^2-1)$}
\end{figure}

\newpage
\section*{Supporting information}
\subsection*{Residence time in a multi-well potential}\label{sec_residence_time_multiwell}
Here we calculate the residence time in the potential of Fig. 
5. As before,  we assume that the potential is zero in the intervals $[0,R] $ and
$[L-R,L]$. The particles start their diffusion from $x=R $. For  an arbitrary distribution of wells widths $p_i$, the potential is equal to $U_0(x/p_i)$ in an interval $x_i<x<x_{i+1} $, where $x_i=\sum_{j<i}p_j$. The mean first passage time, until a particle either reaches to $x=L $ or returns to $x=0$ (with $J_{\text{e}} =0$, is:
\begin{equation}\label{mrt_appendix}
 T(U)/D=P_{tr}(U)T_{\rightarrow}(U)+(1-P_{tr}(U))T_{\leftarrow }(U)
\end{equation}
where
$T_{\rightarrow }(U)=\int_{R}^L dy e^{U(y)}\int_R^y dz e^{-U(z)} $;
$T_{\leftarrow }(U)= \int_{0}^{R} dy e^{U(y)}\int_y^{R} dz e^{-U(z)} $ and
$P_{tr}=\int_{0}^{R}e^{U(x)}dx/\int_0^{L} e^{U(x)} $ is the translocation probability \cite{gardiner}.

Thus, 
$T_{\leftarrow}=R^2/2$ and
\begin{eqnarray}
T_{\rightarrow }(U)=\frac{1}{2}R^2+R\int_{0}^{p_0} e^{-U(x)} dx+\int_0^{p_0} e^{U(y)}dy\int_0^ye^{-U(x)}dx\nonumber
\end{eqnarray}
The second and the third terms in the above expression become, respectively: \begin{eqnarray}
I_1  &=& \sum_{i=1}^n\int_{x_i}^{x_{i+1}}e^{U_0(\frac{x}{p_i})}dx= p_0\int_0^1e^{U_0(u)}du\nonumber\\ I_2 &=& \sum_{i=1}^n\int_0^{p_i}e^{U_0(\frac{y}{p_i})} dy \left[\int_0^y e^{-U_0(\frac{u}{p_i})}du+\sum_{j=1}^{i-1}\int_0^{p_j}e^{-U_0(\frac{u}{p_j})}du
\right]\nonumber\\
&=& (\sum_{i=1}^np_i^2+2\sum_{i=1}^n\sum_{j=1}^{i-1}p_ip_j)I_{0}=p_0^2 I_{0}\nonumber
\end{eqnarray}
where  $I_{0}= \int_0^1 e^{ U_0(x)} dx\int_0^x  e^{-
U_0(y)}dy =\frac{1}{2}\int_0^1 \int_0^1 dx dy e^{U_0(x)}e^{-U_0(y)} $.
Thus, the mean residence time $T(U) $ does not depend on the number of wells $n$. 

The effect of the additional exit current $J_{\text{e}} $ at $x=L-R $ is formally equivalent to shortening the interval $[L-R,L] $ to $[L-R/(1+K),L] $. Repeating the calculation above for this case shows that also for non-zero $J_{\text{e}} $ the residence time does not depend on the number of wells.


\end{document}